\documentclass[aps,pra,notitlepage,10pt,superscriptaddress]{revtex4-2}
\pdfoutput=1
\usepackage[utf8]{inputenc}
\usepackage[english]{babel}
\usepackage[T1]{fontenc}
\usepackage{physics}
\usepackage{mplmacros}
\begin{document}
\title{Entanglement in restricted experimental setups}

\author{Rafiul Nakib}
\email{rafiul.nakib@sydney.edu.au}
\affiliation{School of Computer Science, The University of Sydney, NSW 2006, Australia}
\author{Martin Pl\'{a}vala}
\email{martin.plavala@itp.uni-hannover.de}
\affiliation{Institut für Theoretische Physik, Leibniz Universität Hannover, 30167 Hannover, Germany}

\begin{abstract}
Entanglement is a necessary resource for steering and Bell nonlocality, yet whether a state is separable takes no account of what an experiment can prepare or measure. We define two notions of separability relative to an experiment, one on either side of separability. Inner separability confines the local states to the preparable sets. Outer separability lets the local factors be self-adjoint operators positive on the available effects but not necessarily positive semidefinite. We find exactly when each departs from separability and that outer separability keeps the operational content of separability for the available measurements: outer separable states, entangled or not, admit local hidden variable models for every behaviour these measurements generate and, when the steered party is unrestricted, local hidden state models for every assemblage the steering party's measurements generate. Such hidden state models do not imply outer separability in general, but we identify conditions under which they do.
\end{abstract}

\maketitle

\section{Introduction}
Entanglement is a characteristic feature of quantum theory and a resource for quantum information processing, with quantum cryptography, quantum teleportation and dense coding as canonical uses \cite{horodecki09,guhne09}. Separable states are those that two parties can prepare by local operations and classical communication; all other states are entangled. An experiment fixes what can be done to a quantum system: which states its sources produce and which effects its detectors resolve. The usual notion of separability takes no account of either. A bipartite state is separable if it is a convex combination of product states, and whether it is one is judged against all states and all measurements, available or not. We ask what separability should mean relative to what is available, and find two natural answers, one for each restriction, which lie on opposite sides of separability itself.

We introduce two notions: inner separability, which corresponds to separable states that can be prepared with restricted preparations, and outer separability, which describes states whose statistics appear separable under a restricted set of measurements. We show that every inner separable state is separable and every separable state is outer separable, that there are separable states that are not inner separable as soon as one of the sets of preparable states is a proper subset of its state space, and that there are outer separable states that are entangled as soon as one party's effect set is genuinely restrictive. We then investigate the relation of outer separability with steering and Bell nonlocality: an outer separable state admits a local hidden state model for every steering assemblage generated by measurements with effects in the set of available measurement outcomes $\effect_A$ when the steered party's effect set is unrestricted, and a local hidden variable model for every behaviour generated by measurements with effects in $\effect_A$ and $\effect_B$. The converse fails in general, but we show that it holds when the steered party is unrestricted and the steering party's effect set is generated by finitely many effects that are linearly independent together with the identity, which we call the genericity assumption.

Separability relative to restricted local sets is not new. Ratanje and Virmani construct single-particle state spaces from the duals of the available measurements, with Bell states separable under Pauli measurements as the example, and call a two-particle state separable when it is a convex combination of products of such states \cite{ratanje11}. They call a set of measurements operator compatible when there is an operator for every string of outcomes that returns it with certainty under the Born rule, and state that for tomographically complete operator compatible measurement sets this separability is equivalent to the existence of a local hidden variable model for those measurements \cite{ratanje11}. They also read their rescaled qubit state spaces in two ways, an enlarged Bloch ball as depolarising noise on the measurements and a shrunken one as depolarising noise on the preparations \cite{ratanje11}; the shrunken one is an instance of Definition~\ref{def:insep} with that subset of the Bloch ball as both local sets. Anwar, Jevtic, Rudolph and Virmani define $\mathcal V$-separability for an arbitrary pair of convex sets of local operators and ask for the smallest local sets for which a given entangled state becomes separable, noting that such a decomposition supplies a local hidden variable model for the measurements on which the local sets are positive \cite{anwar15,anwar19}; both of our notions are instances of theirs. Inner separability with one local set the whole state space and the other the states diagonal in a fixed basis is the set of quantum-incoherent states of the resource theory of coherence \cite{chitambar16,streltsov17}, and with local sets cut out of the state spaces by affine constraints it is the constrained separability introduced in \cite{berta21} and so named in \cite{ohst26}. The framework in which outer separability sits is supplied by entanglement relative to a subspace of observables \cite{barnum03,bkov03,barnum05} and by theories whose effect sets are restricted \cite{janottalal13,filippov19,plavala22jpa} and \cite[\S3.7]{plavala22}, which relax the no-restriction hypothesis of \cite{chiribella10}, the assumption that every effect is available.

On the side of hidden state models, joint measurability of a set of measurements is equivalent to their inability to steer any state \cite{uola14,quintino15}. For a class of states, steerability by a given set of measurements is equivalent to the preparation contextuality of a restricted theory built from the state and the measurements \cite{plavala22jpa}. An equivalence between a local hidden state model and a separability condition is known at the level of assemblages \cite{jencova22,plavala17}. For two qubits, necessary and sufficient geometric conditions for steerability are known under projective measurements \cite{nguyenvu16a,nguyen19} and under two-outcome measurements \cite{nguyenvu16b}.

What is new is the pairing of the two notions relative to one experiment, for arbitrary closed convex preparable sets and arbitrary effect sets, whereas Ratanje and Virmani have the two depolarised qubit instances on either side of the quantum case. Also new are the exact conditions under which each notion departs from separability. The genericity assumption under which, for an unrestricted steered party and a steering party whose effect set the effects generate, outer separability is equivalent to the existence of hidden state models for every family of measurements from that effect set is the operator compatibility of Ratanje and Virmani for the list of two-outcome measurements the effects define, without their tomographic completeness.

Section~\ref{sec:prelim} fixes notation and recalls separability, and Section~\ref{sec:defs} defines the two notions. Section~\ref{sec:relation} relates them to separability, Section~\ref{sec:models} shows that outer separable states admit local hidden variable models for the available measurements and, when the steered party is unrestricted, local hidden state models, and Section~\ref{sec:converse} examines when, conversely, such hidden state models imply outer separability. The paper closes with a summary and the questions it leaves open in Section~\ref{sec:conclusions}.

\section{Preliminaries and notation}\label{sec:prelim}

Let $\Ha$ denote a finite-dimensional complex Hilbert space. We will use $\bound(\Ha)$ to denote the set of linear operators on $\Ha$ and $\bound(\Ha)_{\mathrm{sa}}$ for its real subspace of self-adjoint operators. $\dens(\Ha)$ will denote the set of density matrices on $\Ha$. That is, $\dens(\Ha) \subset \bound(\Ha)_{\mathrm{sa}}$ and $\rho \in \dens(\Ha)$ if and only if $\rho \succeq 0$, meaning that all eigenvalues of $\rho$ are nonnegative, and $\Tr(\rho) = 1$. A measurement on a finite-dimensional quantum system is represented by a positive operator-valued measure (POVM), a finite family $\{M_a\}_a \subset \bound(\Ha)_{\mathrm{sa}}$ with $M_a \succeq 0$ and $\sum_a M_a = \I$, and the probability of observing outcome $a$ given the state $\rho$ was prepared is $\Tr(\rho M_a)$. Here $\I$ denotes the identity operator, and for self-adjoint $X$, $Y$ we write $X\preceq Y$ if $Y-X\succeq0$. An effect is an operator $E$ with $0\preceq E\preceq\I$; the members of a POVM are effects.

Throughout, $\Ha_A$ and $\Ha_B$ denote two nonzero finite-dimensional complex Hilbert spaces, of dimensions $d_A$ and $d_B$. For a subset $S\subseteq\bound(\Ha)_{\mathrm{sa}}$, $\conv S$ denotes its convex hull, the set of convex combinations of members of $S$, and the dual cone is $S^{\circ} := \{\xi\in\bound(\Ha)_{\mathrm{sa}} : \Tr(\xi M)\ge 0\text{ for all }M\in S\}$. The dual cone is a closed convex cone; if every member of $S$ is positive semidefinite, then $S^{\circ}$ contains every positive semidefinite operator. For the set of all effects, $S=\{E:0\preceq E\preceq\I\}$, the dual cone is exactly the cone of positive semidefinite operators. The partial traces $\Tr_B:\bound(\Ha_A\otimes\Ha_B)\to\bound(\Ha_A)$ and $\Tr_A:\bound(\Ha_A\otimes\Ha_B)\to\bound(\Ha_B)$ are the linear maps determined by $\Tr_B(X\otimes Y)=\Tr(Y)\,X$ and $\Tr_A(X\otimes Y)=\Tr(X)\,Y$ for all $X\in\bound(\Ha_A)$, $Y\in\bound(\Ha_B)$. On $\mathbb{C}^2$ with its computational basis ${\ket0},{\ket1}$ we write $\sigma_x:=\dyad{0}{1}+\dyad{1}{0}$, $\sigma_y:=-i\dyad{0}{1}+i\dyad{1}{0}$ and $\sigma_z:=\dyad{0}-\dyad{1}$ for the Pauli operators. For an orthonormal basis $\{u_k\}$ of $\bound(\Ha_A)_{\mathrm{sa}}$, every self-adjoint $X$ on $\Ha_A\otimes\Ha_B$ is $\sum_ku_k\otimes Y_k$ with self-adjoint $Y_k=\Tr_A[(u_k\otimes\I_B)X]$, since the products of the $u_k$ with the members of an orthonormal basis of $\bound(\Ha_B)_{\mathrm{sa}}$ form an orthonormal basis of $\bound(\Ha_A\otimes\Ha_B)_{\mathrm{sa}}$.

Consider a state $\rho_{AB} \in \dens(\Ha_A \otimes \Ha_B)$. Then we say that $\rho_{AB}$ is separable \cite{werner89,horodecki09} if there are $\sigma^A_\lambda \in \dens(\Ha_A)$, $\sigma^B_\lambda \in \dens(\Ha_B)$, and probabilities $p(\lambda)$, $p(\lambda) \geq 0$ and $\sum_\lambda p(\lambda) = 1$, such that $\rho_{AB} = \sum_\lambda p(\lambda) \sigma^A_\lambda \otimes \sigma^B_\lambda$. If $\rho_{AB}$ is not separable, then it is termed entangled. It is well known that the set of separable states coincides with the set of states that can be prepared via local operations and classical communication (LOCC) \cite[\S XII.G]{horodecki09}.

\section{Inner and outer separability}\label{sec:defs}

For a quantum system described by a finite-dimensional complex Hilbert space $\Ha$, we will use $K$ to denote the subset of preparable states and $\effect$ the set of observable measurement outcomes. That is, $K \subseteq \dens(\Ha)$ is the closed convex set of density matrices that we can prepare in a given experiment and $\effect \subseteq \{E: 0 \preceq E \preceq \I\}$ is the convex set of measurement effects that we can use to construct POVMs. We call $\effect$ the effect set of the system; it is unrestricted when it is the whole of $\{E:0\preceq E\preceq\I\}$, and restricted otherwise. We require $\I\in\effect$: the outcome that always occurs is among the observable ones. Given two quantum systems corresponding to the Hilbert spaces $\Ha_A$ and $\Ha_B$ we get the respective sets of preparable states $K_A$ and $K_B$ and the sets of observable measurement outcomes $\effect_A$ and $\effect_B$. We can then introduce the following two definitions:
\begin{definition}[Inner separability]\label{def:insep}
A quantum state $\rho_{AB} \in \dens(\Ha_A \otimes \Ha_B)$ is called inner separable if there are probabilities $p(\lambda)$, $p(\lambda) \geq 0$ and $\sum_\lambda p(\lambda) = 1$, and $\omega^A_\lambda \in K_A$, $\omega^B_\lambda \in K_B$ such that
\begin{equation}
\rho_{AB} = \sum_\lambda p(\lambda) \omega^A_\lambda \otimes \omega^B_\lambda.
\end{equation}
\end{definition}

The intuition behind inner separability is that it corresponds to states preparable by LOCC protocols in which every local state either party holds lies in $K_A$ and $K_B$ respectively. The name inner separability captures the fact that the set of inner separable states is a subset of the set of separable states. In the terms of the literature, the inner separable states form the minimal tensor product of $K_A$ and $K_B$, that is, the convex hull of the products $\omega^A\otimes\omega^B$ with $\omega^A\in K_A$ and $\omega^B\in K_B$ \cite[Def.~5.3]{plavala22}.

\begin{definition}[Outer separability]\label{def:outsep}
A quantum state $\rho_{AB} \in \dens(\Ha_A \otimes \Ha_B)$ is called outer separable if there are $\xi^A_\lambda \in \bound(\Ha_A)_{\mathrm{sa}}$, $\xi^B_\lambda \in \bound(\Ha_B)_{\mathrm{sa}}$ such that $\Tr(\xi^A_\lambda E_A) \geq 0$, $\Tr(\xi^B_\lambda E_B) \geq 0$ for all $\lambda$, $E_A \in \effect_A$, $E_B \in \effect_B$, and
\begin{equation}
\rho_{AB} = \sum_\lambda \xi^A_\lambda \otimes \xi^B_\lambda.
\end{equation}
\end{definition}

One can see outer separability as the dual of inner separability: where inner separability asks the local factors to lie in the preparable sets, outer separability asks them to lie in the dual cones of the effect sets. The name outer separability is due to the fact that the set of outer separable states is a superset of the set of separable states.

The operators $\xi^A_\lambda$ and $\xi^B_\lambda$ in a sum of products of this form are called its $A$- and $B$-factors; in a decomposition as in Definition~\ref{def:outsep} they lie in the dual cones $\effect_A^\circ$ and $\effect_B^\circ$. Since $\I\in\effect$, every $\xi\in\effect^\circ$ has $\Tr\xi=\Tr(\xi\I)\ge0$, by the dual-cone condition at $\I$; so the factors of a decomposition as in Definition~\ref{def:outsep} have nonnegative trace.

Definition~\ref{def:outsep} imposes no normalisation on the factors, and the probabilities of Definition~\ref{def:insep} have no counterpart in it: the dual cones are cones, so positive scalars move freely between the factors and any weights would be absorbed. A decomposition whose factors all have positive trace can be written with trace-one factors and probabilities, $\rho_{AB}=\sum_\lambda p(\lambda)\,(\xi^A_\lambda/\Tr\xi^A_\lambda)\otimes(\xi^B_\lambda/\Tr\xi^B_\lambda)$ with $p(\lambda):=\Tr(\xi^A_\lambda)\Tr(\xi^B_\lambda)$, which sum to $\Tr\rho_{AB}=1$, as in \cite{anwar15}; only factors of trace zero escape this form. Such a factor gives zero to every member of every POVM from its effect set, the members' values being nonnegative and summing to the trace, so its products contribute nothing to the statistics of the available measurements; this is the observation of \cite{anwar15,anwar19} on traceless terms in a $\mathcal V$-separable decomposition. Factors of trace zero must nevertheless be admitted by the definition, for two reasons. As it stands, the set of outer separable states is closed: in finite dimension the finite sums of products of elements of two closed convex cones form a closed cone \cite[Thm.~6.10]{vandobbendebruyn20}, and the outer separable states are the states in this cone, its intersection with the closed set $\dens(\Ha_A\otimes\Ha_B)$. With trace-one factors and probabilities it would not be closed in general, and it would in general miss states that the statistics of the available measurements cannot tell from outer separable ones.

To demonstrate why we require factors with zero trace, take $\effect_A=\conv\{0,\I_A,\dyad0,\dyad1\}$ on $\mathbb{C}^2$, $\effect_B$ unrestricted, and $\dyad{\Phi^+}$ with $\ket{\Phi^+}=(\ket{00}+\ket{11})/\sqrt2$. It is outer separable:
\begin{equation}
\dyad{\Phi^+}=\tfrac12 \dyad0\otimes\dyad0 + \tfrac12 \dyad1\otimes\dyad1+ \tfrac18 \sum_{s=\pm1}\bigl[s\sigma_x\otimes(\I_B+s\sigma_x)-s\sigma_y\otimes(\I_B+s\sigma_y)\bigr] ,
\end{equation}
the sum over $s$ being $\tfrac12(\dyad{00}{11}+\dyad{11}{00})$, and $\pm\sigma_x$, $\pm\sigma_y$ lie in $\effect_A^\circ$, having zero trace against $\dyad0$, $\dyad1$ and $\I_A$; these four factors have trace zero. Yet $\dyad{\Phi^+}$ has no decomposition with trace-one factors and probabilities. In such a decomposition $\sum_\lambda p(\lambda)\,\xi^A_\lambda\otimes\xi^B_\lambda$ the $B$-factors are density matrices, $\effect_B$ being unrestricted, and $\Tr_A[(\dyad0\otimes\I_B)\dyad{\Phi^+}]=\tfrac12\dyad0$ and $\Tr_A[(\dyad1\otimes\I_B)\dyad{\Phi^+}]=\tfrac12\dyad1$ are the sums of the positive semidefinite terms $p(\lambda)\Tr(\xi^A_\lambda\dyad0)\,\xi^B_\lambda$ and $p(\lambda)\Tr(\xi^A_\lambda\dyad1)\,\xi^B_\lambda$; the two traces are nonnegative with sum one, so every $\xi^B_\lambda$ of positive weight is a multiple of $\dyad0$ or $\dyad1$, and the entry $\sum_\lambda p(\lambda)\bra0\xi^A_\lambda\ket1\bra0\xi^B_\lambda\ket1$ of the decomposition between $\ket{00}$ and $\ket{11}$ vanishes, whereas $\braket{00}{\Phi^+}\!\!\braket{\Phi^+}{11}=\tfrac12$. It is nevertheless a limit of states that have one: adding $\varepsilon\I_A$ to each of the four factors $\pm\sigma_x$, $\pm\sigma_y$ keeps them in $\effect_A^\circ$ and adds $\tfrac\varepsilon2\I_A\otimes\I_B$ to the sum, so $(\dyad{\Phi^+}+\tfrac\varepsilon2\I_A\otimes\I_B)/(1+2\varepsilon)$, $\varepsilon>0$, has a decomposition whose factors all have positive trace, hence one with trace-one factors and probabilities, and it tends to $\dyad{\Phi^+}$ as $\varepsilon\to0$. And $\dyad{\Phi^+}$ gives the same statistics as the separable state $\tfrac12(\dyad0\otimes\dyad0+\dyad1\otimes\dyad1)$ for every POVM from $\effect_A$ and every POVM of the other party.

We write $\mathrm{OutSep}(\effect_A,\effect_B)$ for the set of outer separable states, that is, of the states that are sums of products $\xi^A\otimes\xi^B$ with $\xi^A\in\effect_A^\circ$ and $\xi^B\in\effect_B^\circ$; this set is convex and, by the paragraphs above, closed. Where $(\effect_A,\effect_B)$ is clear from context we drop the arguments and write simply $\mathrm{OutSep}$.

\section{Relation to separability}\label{sec:relation}

This section places inner and outer separability against separability itself: first the two inclusions, then when each of them is an equality, and finally when outer separability is strictly weaker than separability. The first relations are direct from the definitions.

\begin{proposition}\label{prop:inclusions}
Any inner separable state is separable, and any separable state is outer separable.
\end{proposition}

\begin{proof}
An inner separable decomposition is itself a separable decomposition, since $K_A\subseteq\dens(\Ha_A)$ and $K_B\subseteq\dens(\Ha_B)$. A separable decomposition $\rho_{AB}=\sum_\lambda p(\lambda)\,\sigma^A_\lambda\otimes\sigma^B_\lambda$ in turn satisfies Definition~\ref{def:outsep} with $\xi^A_\lambda:=p(\lambda)\,\sigma^A_\lambda$, $\xi^B_\lambda:=\sigma^B_\lambda$: $\Tr(\sigma E)\ge0$ for every density matrix $\sigma$ and every effect $E$, both being positive semidefinite, and $p(\lambda)\ge0$.
\end{proof}

\begin{proposition}\label{prop:insep-vs-sep}
Let $K_A,K_B$ be closed and convex. If $K_A = \dens(\Ha_A)$ and $K_B = \dens(\Ha_B)$, then $\rho_{AB} \in \dens(\Ha_A \otimes \Ha_B)$ is separable if and only if it is inner separable. If $K_A \subsetneq \dens(\Ha_A)$ or $K_B \subsetneq \dens(\Ha_B)$ then there are states that are separable but not inner separable.
\end{proposition}
\begin{proof}
When $K_A=\dens(\Ha_A)$ and $K_B=\dens(\Ha_B)$, any separable decomposition $\rho_{AB}=\sum_{\lambda} p(\lambda)\,\sigma^{A}_{\lambda}\otimes\sigma^{B}_{\lambda}$ with $\sigma^{A}_{\lambda}\in\dens(\Ha_{A})$, $\sigma^{B}_{\lambda}\in\dens(\Ha_{B})$ automatically satisfies Definition~\ref{def:insep} since all local terms belong to $K_A$ and $K_B$; conversely every inner separable state is separable by Proposition~\ref{prop:inclusions}, so separability and inner separability coincide.

For the strict inclusion, assume without loss of generality that $K_A\subsetneq\dens(\Ha_A)$, choose $\tau\in\dens(\Ha_A)\setminus K_A$, and let $\rho_{AB}:=\tau\otimes\sigma_B$ for any $\sigma_B\in\dens(\Ha_B)$; it is separable. If it were inner separable, with a decomposition $\rho_{AB}=\sum_{\lambda} p(\lambda)\,\omega^{A}_{\lambda}\otimes\omega^{B}_{\lambda}$ as in Definition~\ref{def:insep}, then taking the partial trace over $B$, each $\omega^{B}_{\lambda}$ having unit trace, would give $\tau=\Tr_B\rho_{AB}=\sum_{\lambda} p(\lambda)\,\omega^{A}_{\lambda}$, a convex combination of members of $K_A$ and hence, $K_A$ being convex, a member of $K_A$; but $\tau\notin K_A$.
\end{proof}

\begin{remark}
As a concrete witness, take $K_A$ to consist of the computational-basis-diagonal states on $\mathbb{C}^2$ and $\tau=\dyad{+}$ with $\ket{+}=(\ket{0}+\ket{1})/\sqrt{2}$, a density matrix outside the closed convex set $K_A$. Then $\tau\otimes\sigma_B$ is separable for any $\sigma_B\in\dens(\Ha_B)$ but not inner separable. For a closed convex $K_A$ the failure is also witnessed by an observable: strict separation of $\tau$ from $K_A$ \cite[\S2.5.1, Example~2.20]{convexseparation} gives $F\in\bound(\Ha_A)_{\mathrm{sa}}$ and $c\in\mathbb{R}$ with $\Tr(F\tau)<c\le\Tr(F\omega)$ for all $\omega\in K_A$ (for the $K_A$ of this remark, $F=-\sigma_x$ and $c=0$), so $F\otimes\I_B$ has expectation below $c$ on $\tau\otimes\sigma_B$ and at least $c$ on every inner separable state.
\end{remark}

\begin{proposition}\label{prop:outsep-unrestricted}
Let $\effect_A = \{ E_A: 0 \preceq E_A \preceq \I_A \}$ and $\effect_B = \{ E_B: 0 \preceq E_B \preceq \I_B \}$. Then $\rho_{AB} \in \dens(\Ha_A \otimes \Ha_B)$ is separable if and only if it is outer separable.
\end{proposition}
\begin{proof}
Separable states are outer separable by Proposition~\ref{prop:inclusions}. Conversely, both effect sets being unrestricted, both dual cones are the cone of positive semidefinite operators; so the factors of a decomposition as in Definition~\ref{def:outsep} are positive semidefinite; dropping the products in which a factor vanishes, dividing each remaining factor by its trace, which is positive, and taking the products of the traces, which sum to $\Tr\rho_{AB}=1$, as the probabilities gives a separable decomposition.
\end{proof}

The next proposition shows that whenever one party's effect set is \emph{genuinely restrictive} --- its dual cone strictly exceeds the positive semidefinite cone --- and the other party's system has dimension at least two, outer separability becomes strictly weaker than separability.

\begin{proposition}\label{prop:outsep-strict}
Let $\effect_A\subseteq\{E_A:0\preceq E_A\preceq\I_A\}$ and $\effect_B\subseteq\{E_B:0\preceq E_B\preceq\I_B\}$ be closed and convex, each containing the identity. If one party's effect set is genuinely restrictive, in the sense that its dual cone strictly exceeds $\{\xi:\xi\succeq 0\}$, and the other party's space has dimension at least two, then there exist outer separable states that are not separable.
\end{proposition}
\begin{proof}
A genuinely restrictive party, say $B$, accepts an operator that is no state: one with a negative eigenvalue, which goes unnoticed because the measurement that would show it is not available. We tidy such an operator into one with a single negative eigenvalue that $B$ still accepts (Step~1). We pair it with a pure state of the other party and add four products of pure states, chosen so that they cancel the negative entry exactly; the sum, divided by its trace, is a state, and it is outer separable by construction, every factor being accepted by its party (Step~2). The cancellation leaves an exact zero on the diagonal, while the four products bring a coupling between the two particles, and that zero gives the entanglement away: the partial transpose, a transpose on $B$'s side alone, moves the coupling into the zero's row, where a zero diagonal entry beside a nonzero entry forces a negative eigenvalue, and the partial transpose of a separable state never has one, by Peres' criterion (Step~3).

\emph{Step 1: the operator $\xi_\star$.} Suppose $\effect_B^\circ\supsetneq\{\xi:\xi\succeq0\}$, so that $d_A\ge2$ by hypothesis; the other case follows by exchanging the roles of $A$ and $B$, under which Definition~\ref{def:outsep} and the definition of separability are symmetric. Both dual cones are convex cones containing every positive semidefinite operator, the members of an effect set being positive semidefinite. Pick $\zeta\in\effect_B^\circ$ with a negative eigenvalue, and a unit eigenvector $\ket1_B$ for it. For $t\ge0$ the operator $\zeta+t(\I_B-\dyad1_B)$ lies in $\effect_B^\circ$, being the sum of $\zeta\in\effect_B^\circ$ and a nonnegative multiple of the projector $\I_B-\dyad1_B\succeq0$ onto $\ket1_B^\perp$. As $\zeta$ is self-adjoint, $\ket1_B^\perp$ is invariant under $\zeta$; so this operator keeps the negative eigenvalue at $\ket1_B$, while its eigenvalues on $\ket1_B^\perp$ are those of $\zeta$ there, raised by $t$. For $t$ large enough these are nonnegative and the trace, $\Tr(\zeta)+t(d_B-1)$, is positive, as $d_B\ge2$: $\I_B\in\effect_B$ gives $\Tr\zeta\ge0$, so $\zeta$ has a positive eigenvalue besides the negative one. Dividing by the trace, a positive number, gives $\xi_\star\in\effect_B^\circ$ with $\Tr(\xi_\star)=1$ and, in an orthonormal eigenbasis $\{\ket j_B\}_{j=0}^{d_B-1}$ containing $\ket1_B$, $\xi_\star=-\kappa\,\dyad1_B+\sum_{j\neq1}\gamma_j\,\dyad j_B$ with $\kappa>0$, $\gamma_j\ge0$.

\emph{Step 2: an outer separable state.} Let $V_A:=\mathrm{span}\{\ket0_A,\ket1_A\}$ for any orthonormal pair in $\Ha_A$ (here $d_A\ge2$ is used) and $V_B:=\mathrm{span}\{\ket0_B,\ket1_B\}$, with $\ket0_B$ the eigenvector of $\xi_\star$ labelled $0$ in that eigenbasis, and on each let $\sigma_x:=\dyad{0}{1}+\dyad{1}{0}$ and $\sigma_y:=-i\dyad{0}{1}+i\dyad{1}{0}$, extended by $0$ on the orthogonal complement, with eigenvectors $\ket{\pm x}=\tfrac1{\sqrt2}(\ket0\pm\ket1)$ and $\ket{\pm y}=\tfrac1{\sqrt2}(\ket0\pm i\ket1)$. Write $\I_{V_A}:=\dyad0_A+\dyad1_A$ and $\I_{V_B}:=\dyad0_B+\dyad1_B$ for the projectors onto $V_A$ and $V_B$, $\I_V$ for either, and $\ket{ij}$ for $\ket i_A\otimes\ket j_B$. From $\dyad{\pm x}=\tfrac12(\I_V\pm \sigma_x)$, $\dyad{\pm y}=\tfrac12(\I_V\pm \sigma_y)$ and $\sigma_x\otimes \sigma_x+\sigma_y\otimes \sigma_y=2(\dyad{01}{10}+\dyad{10}{01})$,
\begin{equation}
\begin{aligned}
Q&:=\sum_{s=\pm1}\Bigl(\dyad{sx}_A\otimes\dyad{-sx}_B+\dyad{sy}_A\otimes\dyad{-sy}_B\Bigr)\\
&\;=\I_{V_A}\otimes\I_{V_B}-\bigl(\dyad{01}{10}+\dyad{10}{01}\bigr).
\end{aligned}
\end{equation}
Set $\rho_{AB}:=(\kappa\,Q+\dyad1_A\otimes\xi_\star)/(4\kappa+1)$. This is a sum of five products with positive coefficients, whose $A$-factors are pure states and whose $B$-factors are pure states or $\xi_\star$; all lie in the respective dual cones, the pure states because they are positive semidefinite and $\xi_\star$ by construction. In the product basis formed from an orthonormal basis of $\Ha_A$ containing $\ket0_A,\ket1_A$ and the eigenbasis $\{\ket j_B\}$, $(4\kappa+1)\rho_{AB}$ has diagonal entries $\kappa,\ \kappa,\ \kappa+\gamma_0,\ 0$ at ${\ket{00}},{\ket{01}},{\ket{10}},{\ket{11}}$ and $\gamma_j$ at $\ket{1j}$ for $j\ge2$, zero elsewhere, and the single off-diagonal pair $-\kappa$ between $\ket{01}$ and $\ket{10}$. It is therefore the direct sum of $\bigl(\begin{smallmatrix}\kappa&-\kappa\\-\kappa&\kappa+\gamma_0\end{smallmatrix}\bigr)$ on $\mathrm{span}\{{\ket{01}},\ket{10}\}$, whose diagonal is positive and whose determinant $\kappa\gamma_0$ is nonnegative, and nonnegative scalars; so $\rho_{AB}\succeq0$, and $\Tr(\rho_{AB})=(4\kappa+\Tr\xi_\star)/(4\kappa+1)=1$. Thus $\rho_{AB}$ is a state with a decomposition as in Definition~\ref{def:outsep}: it is outer separable.

\emph{Step 3: the state is not separable.} Transposition on $B$ in the basis $\{\ket j_B\}$, written $\rho\mapsto\rho^{\mathsf T_B}$, fixes every diagonal entry and carries $\dyad{01}{10}$ to $\dyad{00}{11}$ and $\dyad{10}{01}$ to $\dyad{11}{00}$. Hence $(4\kappa+1)\rho_{AB}^{\mathsf T_B}$ is the direct sum of $\bigl(\begin{smallmatrix}\kappa&-\kappa\\-\kappa&0\end{smallmatrix}\bigr)$ on $\mathrm{span}\{{\ket{00}},\ket{11}\}$, of determinant $-\kappa^2<0$, and nonnegative scalars, and so has a negative eigenvalue. The partial transpose of a separable state, $\sum_\lambda p(\lambda)\,\sigma^A_\lambda\otimes(\sigma^B_\lambda)^{\mathsf T}$, is positive semidefinite (Peres' criterion \cite{peres96}); therefore $\rho_{AB}$ is not separable.
\end{proof}

The criterion that either of the effect sets is genuinely restrictive concerns the dual cones, not the effect sets. A genuinely restrictive effect set is restricted, that is, it violates the no-restriction hypothesis \cite{chiribella10}; the converse fails. On a system of dimension at least two, $\effect:=\conv\{F,\I-F:0\preceq F\preceq\I/2\}$ contains $0$ and $\I$ and the complement of each of its members, but no rank-one projector: a member is a convex combination of generators, with total weight $s\in[0,1]$ on those of the form $F$, so it lies between $\tfrac{1-s}{2}\I$ and $(1-\tfrac s2)\I$ and cannot have both eigenvalues $0$ and $1$. Yet its dual cone is exactly the positive semidefinite cone, since the dual cone depends only on the cone generated by the effect set, which here is the cone of all positive semidefinite operators.

\section{Steering and Bell nonlocality}\label{sec:models}
Separable states have local hidden state models \cite{wiseman07,cavalcanti16} and local hidden variable models \cite{werner89,brunner14}, so entanglement is a necessary resource for both steering and Bell nonlocality. Here we show that outer separable states, too, have such models, for the assemblages and behaviours generated by the available measurements.

\subsection{Steering}
Let $\{\sigma_{a|x}\}_{a,x}$ be a family of operators on $\Ha_B$ indexed by measurement labels $x$ and outcomes $a$, an assemblage. The party holding $\Ha_A$, Alice, measures; the party holding $\Ha_B$, Bob, is the steered party. A local hidden state model \cite{wiseman07,cavalcanti16,uola15} for it consists of a probability measure $\mu$ on a measurable space $\Lambda$, for each $x$ and each $\lambda\in\Lambda$ a probability distribution $p(\cdot|x,\lambda)$ over the outcomes of $x$, measurable in $\lambda$, and density matrices $\sigma_\lambda\in\dens(\Ha_B)$, measurable in $\lambda$, such that
\begin{equation}
\sigma_{a|x}=\int_\Lambda p(a|x,\lambda)\,\sigma_\lambda\,\mu(d\lambda)
\end{equation}
for all $a,x$; one and the same $(\Lambda,\mu,\sigma_\lambda)$ serves the whole family, only the responses depending on $x$.

\begin{proposition}\label{prop:osep-implies-lhs}
Let $\rho_{AB} \in \dens(\Ha_A \otimes \Ha_B)$ be a bipartite quantum state, let $\effect_A$, $\effect_B$ be the respective sets of observable measurement outcomes, with $\effect_B = \{E: 0 \preceq E \preceq \I_B \}$ unrestricted. Consider POVMs $\{ M_{a|x} \}_{a,x} \subset \bound(\Ha_A)_{\mathrm{sa}}$ where $a$ labels measurement outcomes and $x$ labels different POVMs, with $M_{a|x} \in \effect_A$. If $\rho_{AB}$ is outer separable, then the steering assemblage $\sigma^B_{a|x}:=\Tr_A[(M_{a|x}\otimes\I_B)\rho_{AB}]$ admits a local hidden state model.
\end{proposition}
\begin{proof}
Let $\rho_{AB}=\sum_{\lambda}\xi^{A}_{\lambda}\otimes\xi^{B}_{\lambda}$ be outer separable. Since $\effect_{B}=\{E:0\preceq E\preceq\I_B\}$ is unrestricted, $\effect_B^\circ$ is the cone of positive semidefinite operators, so each $\xi^B_\lambda$ is positive semidefinite; and $\Tr\xi^A_\lambda\ge0$, as $\I_A\in\effect_A$. For Alice's POVMs $\{M_{a|x}\}_{a,x}\subset\effect_{A}$ the numbers $\Tr(\xi^{A}_{\lambda}M_{a|x})$ are nonnegative, by outer separability, and sum over $a$ to $\Tr\xi^A_\lambda$; so the products in which $\Tr\xi^A_\lambda=0$ or $\xi^B_\lambda=0$ contribute nothing to the assemblage $\sigma^B_{a|x}=\sum_\lambda\Tr(\xi^A_\lambda M_{a|x})\,\xi^B_\lambda$ and are dropped.

For the remaining $\lambda$ put $p(\lambda):=\Tr(\xi^A_\lambda)\Tr(\xi^B_\lambda)>0$, $p(a|x,\lambda):=\Tr(\xi^{A}_{\lambda}M_{a|x})/\Tr\xi^A_\lambda$ and $\sigma_\lambda:=\xi^B_\lambda/\Tr\xi^B_\lambda\in\dens(\Ha_B)$. Then $\sum_\lambda p(\lambda)=\Tr\rho_{AB}=1$, the dropped products having trace zero, each $p(\cdot|x,\lambda)$ is a probability distribution over the outcomes of $x$, and
\begin{equation}
\sigma^{B}_{a|x}=\sum_{\lambda}p(\lambda)\,p(a|x,\lambda)\,\sigma_{\lambda} .
\end{equation}
\end{proof}

\subsection{Bell nonlocality}
Let $\{P(a,b|x,y)\}$ be a family of numbers indexed by settings $x$, $y$ and outcomes $a$, $b$, a behaviour. A local hidden variable model \cite{bell64,brunner14} for it consists of a probability measure $\mu$ on a measurable space $\Lambda$ and, for each $\lambda\in\Lambda$, probability distributions $p(\cdot|x,\lambda)$ over the outcomes of each $x$ and $p(\cdot|y,\lambda)$ over the outcomes of each $y$, measurable in $\lambda$, such that
\begin{equation}
P(a,b|x,y)=\int_\Lambda p(a|x,\lambda)\,p(b|y,\lambda)\,\mu(d\lambda)    
\end{equation}
for all $a,b,x,y$.

\begin{proposition}\label{prop:osep-implies-lhv}
Let $\rho_{AB} \in \dens(\Ha_A \otimes \Ha_B)$ be a bipartite quantum state and let $\effect_A$, $\effect_B$ be the respective sets of observable measurement outcomes. Consider POVMs $\{ M_{a|x} \}_{a,x} \subset \bound(\Ha_A)_{\mathrm{sa}}$ and $\{ N_{b|y} \}_{b,y} \subset \bound(\Ha_B)_{\mathrm{sa}}$ with $M_{a|x} \in \effect_A$, $N_{b|y} \in \effect_B$. If $\rho_{AB}$ is outer separable, then the behaviour $P_\rho(a,b|x,y):=\Tr[\rho_{AB}(M_{a|x}\otimes N_{b|y})]$ admits a local hidden variable model.
\end{proposition}
\begin{proof}
Let $\rho_{AB}=\sum_{\lambda}\xi^{A}_{\lambda}\otimes\xi^{B}_{\lambda}$ be outer separable. As $\I_A\in\effect_A$ and $\I_B\in\effect_B$, the traces $\Tr\xi^A_\lambda$ and $\Tr\xi^B_\lambda$ are nonnegative. The numbers $\Tr(\xi^{A}_{\lambda}M_{a|x})$ are nonnegative, by outer separability, and sum over $a$ to $\Tr\xi^A_\lambda$, and likewise the $\Tr(\xi^{B}_{\lambda}N_{b|y})$ on Bob's side; so the products in which a factor has trace zero contribute nothing to $P_\rho$ and are dropped. For the remaining $\lambda$ put $p(\lambda):=\Tr(\xi^A_\lambda)\Tr(\xi^B_\lambda)>0$; these sum to $\Tr\rho_{AB}=1$. The joint behaviour factorises:
\begin{equation}
P_\rho(a,b|x,y)=\sum_{\lambda}p(\lambda)\,p(a|x,\lambda)\,p(b|y,\lambda),
\end{equation}
with $p(a|x,\lambda):=\Tr(\xi^{A}_{\lambda}M_{a|x})/\Tr\xi^A_\lambda$ and $p(b|y,\lambda):=\Tr(\xi^{B}_{\lambda}N_{b|y})/\Tr\xi^B_\lambda$, probability distributions over the outcomes of $x$ and of $y$.
\end{proof}

\section{When does the converse hold?}\label{sec:converse}
Proposition~\ref{prop:osep-implies-lhs} has no converse in general, just as entanglement, steering and Bell nonlocality are inequivalent for general measurements~\cite{quintino15ineq}.

\subsection{The converse fails in general}

\begin{proposition}\label{prop:converse-fails}
There exist bipartite states $\rho_{AB}$ and restricted effect sets $\effect_A$ such that $\rho_{AB}$ admits a local hidden state model for all steering assemblages generated by POVMs with effects in $\effect_A$, but $\rho_{AB}$ is not outer separable with respect to $(\effect_A, \effect_B)$ where $\effect_B=\{E:0\preceq E\preceq\I_B\}$.
\end{proposition}

\begin{proof}
The effect set is the image of the set of all qubit effects under the depolarising map at visibility one half. Every family of POVMs with effects in it is then jointly measurable, which supplies the models; and, for the maximally entangled state, outer separability would make its depolarised image separable, which it is not.

Consider qubits $\Ha_A=\Ha_B=\mathbb{C}^2$ and define the depolarising channel $\Delta_\eta(X):=\eta X+(1-\eta)\Tr(X)\,\I/2$. Let $\effect_A:=\{\Delta_{1/2}(E):0\preceq E\preceq\I_A\}$ be the image of the set of all effects under $\Delta_{1/2}$, and let $\effect_B:=\{E:0\preceq E\preceq\I_B\}$. The set $\effect_A$ is a restricted effect set: it is convex, as the image of a convex set under a linear map; it contains $\I_A=\Delta_{1/2}(\I_A)$, as every effect set must; its members are effects, as $\Delta_{1/2}(E)\succeq0$ and $\I_A-\Delta_{1/2}(E)=\Delta_{1/2}(\I_A-E)\succeq0$ for $0\preceq E\preceq\I_A$; and it is not the set of all effects, since the only preimage of $\dyad0$ under the invertible map $\Delta_{1/2}$ is $\mathrm{diag}(\tfrac32,-\tfrac12)$, which is not an effect. Since $\Delta_{1/2}$ is invertible and unital, any POVM with effects in $\effect_A$ is the $\Delta_{1/2}$-image of a POVM: $\sum_a\Delta_{1/2}(E_a)=\I_A=\Delta_{1/2}(\I_A)$ forces $\sum_aE_a=\I_A$. By \cite{renner24, zhang24}, at visibility $\eta=\tfrac12$ every family of qubit POVMs with effects in $\effect_A$ is therefore jointly measurable, that is, obtained from a single parent measurement by post-processing its outcome. For a jointly measurable family with a finite parent, the assemblage of any fixed bipartite state admits a local hidden state model, the outcome of the parent serving as the hidden variable \cite{uola14,quintino15,uola15}. The parent in \cite{renner24} is continuous, a positive operator-valued measure on the unit sphere rather than a finite family; the same argument with an integral in place of the sum \cite{cavalcanti16} gives, for this parent and any fixed bipartite state $\rho_{AB}$, a local hidden state model with a continuous hidden variable for the assemblage built from $\effect_A$. In particular, the maximally entangled state $\rho_{AB}=\dyad{\Phi^+}$ with $\ket{\Phi^+}=(\ket{00}+\ket{11})/\sqrt2$ admits a local hidden state model for all steering assemblages built from $\effect_A$.

Suppose, for contradiction, that $\rho_{AB}$ is outer separable with respect to $(\effect_A,\effect_B)$, with a decomposition $\rho_{AB}=\sum_\lambda\xi^A_\lambda\otimes\xi^B_\lambda$ as in Definition~\ref{def:outsep}; its $B$-factors are positive semidefinite, since $\effect_B$ is unrestricted. The dual-cone condition $\Tr(\xi^A_\lambda\,\Delta_{1/2}(E))\ge0$ for all effects $E$ reads $\Tr(\Delta_{1/2}(\xi^A_\lambda)\,E)\ge0$, since $\Tr(\xi\,\Delta_{1/2}(E))=\tfrac12\Tr(\xi E)+\tfrac14\Tr(\xi)\Tr(E)=\Tr(\Delta_{1/2}(\xi)\,E)$; so $\Delta_{1/2}(\xi^A_\lambda)\succeq0$, that is, $\xi^A_\lambda+\tfrac12\Tr(\xi^A_\lambda)\,\I_A\succeq0$. Applying $\Delta_{1/2}\otimes\mathrm{id}$ to the decomposition gives
\begin{equation}
(\Delta_{1/2}\otimes\mathrm{id})(\rho_{AB})=\sum_\lambda\Delta_{1/2}(\xi^A_\lambda)\otimes\xi^B_\lambda,
\end{equation}
a state that is a sum of products of positive semidefinite operators and so, after dropping the vanishing products and dividing the others by the traces of their factors, a convex combination of product density matrices: it is separable. However, $(\Delta_{1/2}\otimes\mathrm{id})(\dyad{\Phi^+})=\tfrac12\dyad{\Phi^+}+\tfrac14\,\I_A\otimes\Tr_A\dyad{\Phi^+}=\tfrac12\dyad{\Phi^+}+\tfrac18\,\I_A\otimes\I_B$. Its partial transpose is the direct sum of the scalars $\tfrac38$ at $\ket{00}$ and $\ket{11}$ and of $\bigl(\begin{smallmatrix}1/8&1/4\\1/4&1/8\end{smallmatrix}\bigr)$ on $\mathrm{span}\{{\ket{01}},\ket{10}\}$, so it has eigenvalues $\tfrac38$ (multiplicity~3) and $-\tfrac18$, and the state is not separable, by Peres' criterion \cite{peres96} as in the proof of Proposition~\ref{prop:outsep-strict}. This contradicts the separability just derived, so $\rho_{AB}$ is not outer separable with respect to $(\effect_A,\effect_B)$; as it admits a local hidden state model for every assemblage built from $\effect_A$, the converse of Proposition~\ref{prop:osep-implies-lhs} fails for this pair.
\end{proof}

\subsection{The converse under the genericity assumption, and for two effects}
Let $N\ge0$ and let $E_1,\ldots,E_N$ be effects on $\Ha_A$. They satisfy the genericity assumption if $\I_A,E_1,\ldots,E_N$ are linearly independent,
\begin{equation}
\dim\mathrm{span}\{\I_A,E_1,\ldots,E_N\}=N+1.
\label{eq:prop4-genericity}
\end{equation}
In words, no $E_i$ is a constant plus a fixed combination of the others, so that no two-outcome measurement $\{E_i,\I_A-E_i\}$ has its outcome probabilities fixed by those of the others for every state. Equivalently, it is the operator compatibility of \cite{ratanje11} for these $N$ measurements, taken as a list in which a repeated measurement counts twice. The genericity assumption is not only sufficient for the existence, for every string $a\in\{0,1\}^N$, of a trace-one operator $\xi_a$ with $\Tr(\xi_aE_i)=1-a_i$ (for instance $\theta_0+\sum_{i=1}^N(1-a_i)\theta_i$, with the dual basis $\theta_0,\ldots,\theta_N$ of Step~1 of the proof of Proposition~\ref{prop:osep-lhs-equiv}) but necessary: without it, some list of values $(1,1-a_1,\ldots,1-a_N)$ is attained by no $\xi$. For if every list were attained, pairing a dependence $c_0\I_A+\sum_ic_iE_i=0$ with the $\xi$ of the list of $a$ would give $c_0+\sum_ic_i(1-a_i)=0$ for every $a$; comparing $a=0$ with the string whose only entry $1$ is $a_j$ gives $c_j=0$ for each $j$, and then $c_0=0$. The existence of a $\xi$ for every list is the operator compatibility of Ratanje and Virmani \cite{ratanje11} for the two-outcome measurements $\{E_i,\I_A-E_i\}$, $i=1,\ldots,N$, taken as a list in which a repeated measurement counts twice; so the genericity assumption is exactly that condition.

\begin{proposition}\label{prop:osep-lhs-equiv}
Let $N\ge0$ and let $E_1,\ldots,E_N\in\bound(\Ha_A)_{\mathrm{sa}}$ be effects, $0\preceq E_i\preceq\I_A$, satisfying the genericity assumption \eqref{eq:prop4-genericity}.
Set $\effect_A:=\conv(\{0,\I_A\}\cup\{E_i,\I_A-E_i\}_{i=1}^N)$ and $\effect_B:=\{E_B:0\preceq E_B\preceq\I_B\}$, and let $\rho_{AB}\in\dens(\Ha_A\otimes\Ha_B)$. Then $\rho_{AB}\in\mathrm{OutSep}(\effect_A,\effect_B)$ if and only if for every family of POVMs $\{M_{a|x}\}_{a,x}\subset\effect_A$ the steering assemblage $\sigma^B_{a|x}:=\Tr_A[(M_{a|x}\otimes\I_B)\rho_{AB}]$ admits a local hidden state model.
\end{proposition}

\begin{proof}
Measurements from $\effect_A$ see $\rho_{AB}$ only through $\rho_B:=\Tr_A\rho_{AB}$ and the operators $\sigma_{E_x}$, $x=1,\ldots,N$, where $\sigma_M:=\Tr_A[(M\otimes\I_B)\rho_{AB}]$ for $M\in\bound(\Ha_A)_{\mathrm{sa}}$. The genericity assumption gives a basis of $\bound(\Ha_A)_{\mathrm{sa}}$ adapted to $\I_A,E_1,\ldots,E_N$, along which $\rho_{AB}$ splits into a part carried by $\rho_B$ and the $\sigma_{E_x}$, which the measurements see, and a part carried by $A$-factors that pair to zero with $\I_A$ and every $E_x$, which they do not. A hidden state model for the $N$ two-outcome measurements $\{E_x,\I_A-E_x\}$ writes the first part as an integral of products of elements of the dual cones, the hidden variable's response probabilities becoming the $A$-factor and its hidden state the $B$-factor, and the closedness of the cone of finite sums of such products makes it a finite sum; the second part becomes such a sum once each of its $B$-coefficients is split into its positive and negative parts, its $A$-factors lying in the dual cone with either sign.

\emph{The forward direction.} If $\rho_{AB}$ is outer separable, Proposition~\ref{prop:osep-implies-lhs} with $\effect_B$ unrestricted gives a local hidden state model for every family of POVMs from $\effect_A$.

\emph{Step 1: the dual basis.} By \eqref{eq:prop4-genericity} the operators $E_0:=\I_A,E_1,\ldots,E_N$ are a basis of their real span $V$, so their Gram matrix $G_{ik}:=\Tr(E_iE_k)$, $i,k=0,\ldots,N$, is positive definite, and $\theta_i:=\sum_{k=0}^N(G^{-1})_{ik}E_k\in V$ satisfy $\Tr(\theta_iE_k)=\delta_{ik}$: $\Tr\theta_0=1$ and $\Tr(\theta_0E_x)=0$, while $\Tr\theta_x=0$ and $\Tr(\theta_xE_y)=\delta_{xy}$ for $x,y\ge1$. The $\theta_i$ are $N+1$ linearly independent elements of $V$, hence a basis of it. Let $\zeta_1,\ldots,\zeta_{n'}$ be an orthonormal basis of the orthogonal complement $V^\perp$ of $V$ in $\bound(\Ha_A)_{\mathrm{sa}}$, $n':=d_A^2-N-1$, empty if $n'=0$. The generators $0,\I_A,E_x,\I_A-E_x$ of $\effect_A$ lie in $V$, so $\effect_A\subseteq V$; hence $\Tr(\zeta_jM)=0$ for every $M\in\effect_A$, and $\pm\zeta_j\in\effect_A^\circ$. Since every $M\in\effect_A$ is a convex combination of the generators, a self-adjoint $\eta$ with $\Tr\eta=1$ lies in $\effect_A^\circ$ if and only if $0\le\Tr(\eta E_x)\le1$ for $x=1,\ldots,N$, the conditions at $0$ and $\I_A$ being void. The genericity assumption enters the proof only through the $\theta_i$.

\emph{Step 2: the expansion of $\rho_{AB}$.} The $d_A^2$ operators $\theta_0,\ldots,\theta_N$, $\zeta_1,\ldots,\zeta_{n'}$ are a basis of $\bound(\Ha_A)_{\mathrm{sa}}$, and $E_0,\ldots,E_N$, $\zeta_1,\ldots,\zeta_{n'}$ is its dual basis for the pairing $\Tr(XY)$: $\Tr(\theta_iE_k)=\delta_{ik}$, $\Tr(\theta_i\zeta_j)=0$ as $\theta_i\in V$, $\Tr(\zeta_jE_k)=0$, $\Tr(\zeta_j\zeta_l)=\delta_{jl}$. Expanding $\rho_{AB}$ along an orthonormal basis of $\bound(\Ha_A)_{\mathrm{sa}}$ and writing each of its members in this basis, with real coefficients, $\rho_{AB}=\sum_i\theta_i\otimes Y_i+\sum_j\zeta_j\otimes Y'_j$ with self-adjoint $Y_i,Y'_j$, and pairing with the dual elements reads the coefficients off, $Y_i=\Tr_A[(E_i\otimes\I_B)\rho_{AB}]$ and $Y'_j=\Tr_A[(\zeta_j\otimes\I_B)\rho_{AB}]$:
\begin{equation}
\rho_{AB}=\theta_0\otimes\rho_B+\sum_{x=1}^N\theta_x\otimes\sigma_{E_x}+\sum_{j=1}^{n'}\zeta_j\otimes\tau_j,\qquad\tau_j:=\Tr_A[(\zeta_j\otimes\I_B)\rho_{AB}] .
\end{equation}

\emph{Step 3: the model.} Now let every family of POVMs from $\effect_A$ have a local hidden state model; in particular the $N$ POVMs $\{E_x,\I_A-E_x\}$ have one, $(\Lambda,\mu,\sigma_\lambda)$ with responses $p(b|x,\lambda)$, the outcome $0$ standing for $E_x$, so that $\sigma_{E_x}=\int_\Lambda p(0|x,\lambda)\,\sigma_\lambda\,\mu(d\lambda)$ and $\rho_B=\sigma_{E_x}+\sigma_{\I_A-E_x}=\int_\Lambda\sigma_\lambda\,\mu(d\lambda)$ (for $N=0$ take $\Lambda$ a point and $\sigma_\lambda=\rho_B$). Put $\eta_\lambda:=\theta_0+\sum_xp(0|x,\lambda)\,\theta_x\in V$. Then $\Tr\eta_\lambda=1$ and $\Tr(\eta_\lambda E_x)=p(0|x,\lambda)\in[0,1]$, so $\eta_\lambda\in\effect_A^\circ$ by Step~1, and
\begin{equation}
\theta_0\otimes\rho_B+\sum_x\theta_x\otimes\sigma_{E_x}=\int_\Lambda\eta_\lambda\otimes\sigma_\lambda\,\mu(d\lambda) .
\end{equation}

\emph{Step 4: finitely many terms.} The finite sums of products of elements of $\effect_A^\circ$ and positive semidefinite operators form a closed convex cone $\mathcal K$ (the paragraphs after Definition~\ref{def:outsep}). An operator $X\notin\mathcal K$ is strictly separated from $\mathcal K$ by a self-adjoint $W$ \cite[\S2.5.1, Example~2.20]{convexseparation}, and since $\mathcal K$ is a cone containing $0$ this gives $\Tr(WY)\ge0$ for all $Y\in\mathcal K$ and $\Tr(WX)<0$. For $X:=\int_\Lambda\eta_\lambda\otimes\sigma_\lambda\,\mu(d\lambda)$ and such a $W$, $\Tr(WX)=\int_\Lambda\Tr(W\,\eta_\lambda\otimes\sigma_\lambda)\,\mu(d\lambda)\ge0$, each $\eta_\lambda\otimes\sigma_\lambda$ lying in $\mathcal K$; so $X\in\mathcal K$:
\begin{equation}
\int_\Lambda\eta_\lambda\otimes\sigma_\lambda\,\mu(d\lambda)=\sum_{k=1}^m\alpha_k\otimes\beta_k,\qquad\alpha_k\in\effect_A^\circ,\ \beta_k\succeq0,
\end{equation}
a finite sum (for $N=0$ it is $\I_A/d_A\otimes\rho_B$).

\emph{Step 5: the part the measurements do not see.} Split $\tau_j=\tau_j^+-\tau_j^-$ into its positive and negative parts, $\tau_j^\pm\succeq0$. Every member of $\effect_B$ being positive semidefinite, $\effect_B^\circ$ contains every positive semidefinite operator, so $\beta_k,\tau_j^\pm\in\effect_B^\circ$, while $\alpha_k,\pm\zeta_j\in\effect_A^\circ$ by Steps~4 and~1. Hence
\begin{equation}
\rho_{AB}=\sum_{k=1}^m\alpha_k\otimes\beta_k+\sum_{j=1}^{n'}\bigl[\zeta_j\otimes\tau_j^++(-\zeta_j)\otimes\tau_j^-\bigr]
\end{equation}
is a decomposition as in Definition~\ref{def:outsep}: $\rho_{AB}\in\mathrm{OutSep}$. The reverse direction used the right-hand side of the statement only through a local hidden state model for the $N$ two-outcome measurements $\{E_x,\I_A-E_x\}$; so $\rho_{AB}$ is outer separable exactly when these $N$ measurements have a local hidden state model, a property of $\rho_B$ and the $\sigma_{E_x}$ alone.

\end{proof}

\begingroup\sloppy
\begin{corollary}\label{cor:two-povm}
Let $E,F$ be two effects, i.e., $0\preceq E\preceq\I_A$ and $0\preceq F\preceq\I_A$, and let $\effect_A=\conv(\{E,\allowbreak F,\allowbreak{\I_A-E},\allowbreak{\I_A-F},\allowbreak 0,\allowbreak\I_A\})$ be the set of measurement outcomes generated by $E$ and $F$. Let $\effect_B=\{E_B:0\preceq E_B\preceq\I_B\}$. Let $\rho_{AB}\in\dens(\Ha_A\otimes\Ha_B)$, and consider POVMs $\{M_{a|x}\}_{a,x}\subset\bound(\Ha_A)_{\mathrm{sa}}$ where $a$ labels measurement outcomes and $x$ labels different POVMs, such that $M_{a|x}\in\effect_A$. Then $\rho_{AB}\in\mathrm{OutSep}(\effect_A,\effect_B)$ if and only if for every such family of POVMs there is a local hidden state model for the steering assemblage $\Tr_A[(M_{a|x}\otimes\I_B)\rho_{AB}]$. No genericity assumption on $E$ and $F$ is required.
\end{corollary}
\endgroup

\begin{proof}
The two effects span with $\I_A$ a space of dimension three, two or one; in the extreme cases the effect set is one of Proposition~\ref{prop:osep-lhs-equiv}'s, at $N=2$ and at $N=0$. In the middle case the effect set of the pair lies in the effect set of a single rescaled effect $E'$, and for a single effect both sides hold at every state: every state has a local hidden state model for the single two-outcome measurement $\{E',\I_A-E'\}$, which gives the left-hand side by the reverse direction of Proposition~\ref{prop:osep-lhs-equiv} at $N=1$, whose proof uses only this measurement, and the left-hand side gives the right-hand side by Proposition~\ref{prop:osep-implies-lhs}.

Let $n:=\dim\mathrm{span}\{\I_A,E,F\}\in\{1,2,3\}$. If $n=3$, the pair $E_1:=E$, $E_2:=F$ satisfies \eqref{eq:prop4-genericity} at $N=2$ and $\effect_A$ is the effect set of Proposition~\ref{prop:osep-lhs-equiv} for it, so the claim is that proposition. If $n=1$, both effects are scalar and $\effect_A=\conv\{0,\I_A\}$, the effect set of Proposition~\ref{prop:osep-lhs-equiv} at $N=0$; the claim is again that proposition. Let now $n=2$, where Proposition~\ref{prop:osep-lhs-equiv} does not apply to the pair. Some member of $\{E,F\}$ is not scalar; call it $G$. Then $\mathrm{span}\{\I_A,G\}=\mathrm{span}\{\I_A,E,F\}$.

Let $g_-<g_+$ be the least and the greatest eigenvalue of $G$, with unit eigenvectors $v_\pm$, and put $E':=(G-g_-\I_A)/(g_+-g_-)$, a non-scalar effect, which satisfies \eqref{eq:prop4-genericity} at $N=1$; let $\effect_A':=\conv\{0,\I_A,E',\I_A-E'\}$ be the effect set of Proposition~\ref{prop:osep-lhs-equiv} for it. Each $M\in\effect_A$ lies in $\mathrm{span}\{\I_A,G\}=\mathrm{span}\{\I_A,E'\}$, so $M=\alpha E'+\beta(\I_A-E')$ with $\alpha=\bra{v_+}M\ket{v_+}$ and $\beta=\bra{v_-}M\ket{v_-}$, both in $[0,1]$ as $M$ is an effect, and $\alpha E'+\beta(\I_A-E')=(1-\alpha)(1-\beta)\cdot0+\alpha\beta\,\I_A+\alpha(1-\beta)\,E'+(1-\alpha)\beta\,(\I_A-E')\in\effect_A'$; so $\effect_A\subseteq\effect_A'$. Hence $(\effect_A')^{\circ}\subseteq\effect_A^{\circ}$, so every decomposition as in Definition~\ref{def:outsep} for $(\effect_A',\effect_B)$ is one for $(\effect_A,\effect_B)$: $\mathrm{OutSep}(\effect_A',\effect_B)\subseteq\mathrm{OutSep}(\effect_A,\effect_B)$.

Every state has a local hidden state model for the single POVM $\{E',\I_A-E'\}$: with $\sigma_M$ and $\rho_B$ as in the proof of Proposition~\ref{prop:osep-lhs-equiv}, $\sigma_{E'}$ and $\sigma_{\I_A-E'}$ are positive semidefinite ($\bra\psi\sigma_M\ket\psi$ is the trace of a product of two positive semidefinite operators) with sum $\rho_B$, so the outcome $b$, drawn with probability $\Tr\sigma_{M_b}$ from the outcomes of positive weight, where $M_0:=E'$ and $M_1:=\I_A-E'$, with the deterministic response $b$ and the hidden state $\sigma_{M_b}/\Tr\sigma_{M_b}$, is such a model, an outcome of weight zero, which the model never returns, having $\sigma_{M_b}=0$, a positive semidefinite operator of trace zero being zero. The reverse direction of Proposition~\ref{prop:osep-lhs-equiv} at $N=1$ for $E'$ uses only this POVM and gives $\rho_{AB}\in\mathrm{OutSep}(\effect_A',\effect_B)\subseteq\mathrm{OutSep}(\effect_A,\effect_B)$; and Proposition~\ref{prop:osep-implies-lhs} then gives a local hidden state model for every family of POVMs from $\effect_A'$, hence from $\effect_A$. Both sides of the equivalence hold for every state, so they are equivalent.
\end{proof}

\subsection{Neither hypothesis can be dropped}

\begin{example}[The genericity assumption cannot be dropped]\label{ex:genericity}
We give six effects on a qubit that violate \eqref{eq:prop4-genericity} and for which, with $\effect_B$ unrestricted as in Proposition~\ref{prop:osep-lhs-equiv}, $\dyad{\Phi^+}$ satisfies the right-hand side of that proposition but not the left-hand side. Let $d_A=d_B=2$, write $\vec n\cdot\vec\sigma:=n_x\sigma_x+n_y\sigma_y+n_z\sigma_z$, and let $E_i:=\tfrac12\I_A+\tfrac14\,\vec n_i\cdot\vec\sigma$ for the six unit vectors $\vec n_i$ along $x$, $y$, $z$, $(x+y)/\sqrt2$, $(x+z)/\sqrt2$ and $(y+z)/\sqrt2$. These are effects, with eigenvalues $\tfrac14$ and $\tfrac34$, and $\mathrm{span}\{\I_A,E_1,\ldots,E_6\}$ has dimension $4<7$.

With $\Delta_{1/2}$ as in the proof of Proposition~\ref{prop:converse-fails} and $P_i:=\tfrac12(\I_A+\vec n_i\cdot\vec\sigma)$, the generators of $\effect_A:=\conv(\{0,\I_A\}\cup\{E_i,\I_A-E_i\}_{i=1}^6)$ are $0=\Delta_{1/2}(0)$, $\I_A=\Delta_{1/2}(\I_A)$, $E_i=\Delta_{1/2}(P_i)$ and $\I_A-E_i=\Delta_{1/2}(\I_A-P_i)$, so $\effect_A$ lies in the effect set of that proof, and $\dyad{\Phi^+}$ has a local hidden state model for every family of POVMs from $\effect_A$: the right-hand side of Proposition~\ref{prop:osep-lhs-equiv} holds.

But $\dyad{\Phi^+}\notin\mathrm{OutSep}$. In a decomposition as in Definition~\ref{def:outsep} the $B$-factors are positive semidefinite, $\effect_B$ being unrestricted, and the $A$-factors have nonnegative trace. An $A$-factor $\xi\in\effect_A^\circ$ of trace zero has $\Tr(\xi E_i)\ge0$ and $\Tr(\xi(\I_A-E_i))=-\Tr(\xi E_i)\ge0$, so $\Tr(\xi E_i)=0$ for every $i$, and as $\I_A,E_1,\ldots,E_6$ span $\bound(\Ha_A)_{\mathrm{sa}}$, $\xi=0$. Dropping the vanishing products and normalising, every $\rho\in\mathrm{OutSep}$ is therefore a convex combination of products $\xi\otimes\tau$ with $\xi\in\effect_A^\circ$ of trace one and $\tau$ a density matrix. A trace-one $\xi\in\effect_A^\circ$ is $\tfrac12(\I_A+\vec r\cdot\vec\sigma)$ with $\Tr(\xi E_i)=\tfrac12+\tfrac14\,\vec r\cdot\vec n_i\in[0,1]$, that is, $|\vec r\cdot\vec n_i|\le2$ for every $i$. On this polytope $|\vec r|^2\le20-8\sqrt2$: two of the coordinates of $\vec r$ share a sign, say $r_j,r_k\ge0$ after $\vec r\mapsto-\vec r$; then $r_j,r_k\le2$ and $r_j+r_k\le2\sqrt2$, and the convex function $r_j^2+r_k^2$ is largest at a vertex of this pentagon, where it is at most $4+(2\sqrt2-2)^2=16-8\sqrt2$, while the third coordinate contributes at most $4$. As $8\sqrt2>11$, this gives $|\vec r|<3$, so the largest eigenvalue $\tfrac12(1+|\vec r|)$ of $\xi$ is at most $\tfrac12\bigl(1+\sqrt{20-8\sqrt2}\bigr)<2$. For every density matrix $\tau$ on $\Ha_B$, $\Tr[\dyad{\Phi^+}(\xi\otimes\tau)]=\tfrac12\Tr(\xi^{\mathsf T}\tau)$, with $\xi^{\mathsf T}$ the transpose in the computational basis, which has the eigenvalues of $\xi$; so $\Tr[\dyad{\Phi^+}(\xi\otimes\tau)]\le\tfrac14\bigl(1+\sqrt{20-8\sqrt2}\bigr)<1$. Hence $\Tr(\dyad{\Phi^+}\rho)$ is at most this bound for every $\rho\in\mathrm{OutSep}$, while it is $1$ at $\rho=\dyad{\Phi^+}$.
\end{example}

\begin{example}[The assumption on the steered party cannot be dropped]\label{ex:steered}
We give a restricted effect set $\effect_B$ for which $\dyad{\Phi^+}$ is outer separable although it has no local hidden state model for three measurements from $\effect_A$. Let $d_A=d_B=2$ and $\effect_B:=\{t\,\I_B:t\in[0,1]\}$, an effect set whose dual cone is $\{\xi\in\bound(\Ha_B)_{\mathrm{sa}}:\Tr\xi\ge0\}$. Let $N=3$ and $E_j:=\tfrac12(\I_A+\sigma_j)$ for $j\in\{x,y,z\}$; these are projectors and hence effects, and they satisfy \eqref{eq:prop4-genericity}, as $\I_A,\sigma_x,\sigma_y,\sigma_z$ are linearly independent. Let $\effect_A$ be the effect set of Proposition~\ref{prop:osep-lhs-equiv} for these three effects, put $s_x=s_z=1$, $s_y=-1$, and let $\rho_{AB}:=\dyad{\Phi^+}$. Then
\begin{equation}
\dyad{\Phi^+}=\sum_{j\in\{x,y,z\}}\tfrac16\Bigl[E_j\otimes\bigl(\tfrac12\I_B+\tfrac32s_j\sigma_j\bigr)+(\I_A-E_j)\otimes\bigl(\tfrac12\I_B-\tfrac32s_j\sigma_j\bigr)\Bigr],
\end{equation}
the $j$-th bracket being $\tfrac12\I_A\otimes\I_B+\tfrac32s_j\,\sigma_j\otimes\sigma_j$ and $\dyad{\Phi^+}=\tfrac14(\I_A\otimes\I_B+\sigma_x\otimes\sigma_x-\sigma_y\otimes\sigma_y+\sigma_z\otimes\sigma_z)$. The $A$-factors are trace-one elements of $\effect_A^\circ$, since $\Tr(E_jE_k)=\tfrac12(1+\delta_{jk})$ and $\Tr((\I_A-E_j)E_k)=\tfrac12(1-\delta_{jk})$ lie in $[0,1]$, and the $B$-factors have trace one, so they lie in $\effect_B^\circ$; the coefficients $\tfrac16$ being absorbed by the cones, $\dyad{\Phi^+}\in\mathrm{OutSep}(\effect_A,\effect_B)$, although the $B$-factors, with eigenvalues $2$ and $-1$, are not positive semidefinite. But the right-hand side of Proposition~\ref{prop:osep-lhs-equiv} fails. A local hidden state model $(\Lambda,\mu,\sigma_\lambda)$ with responses $p(b|j,\lambda)$ for the three POVMs $\{E_j,\I_A-E_j\}$ would give, with $\sigma_M$ and $\rho_B$ as in the proof of Proposition~\ref{prop:osep-lhs-equiv}, $\sigma_{E_j}=\int_\Lambda p(0|j,\lambda)\,\sigma_\lambda\,\mu(d\lambda)$ and $\rho_B=\int_\Lambda\sigma_\lambda\,\mu(d\lambda)$. Here $\rho_B=\tfrac12\I_B$ and $\sigma_{E_j}=\tfrac12E_j^{\mathsf T}=\tfrac14(\I_B+s_j\sigma_j)$, as $\sigma_y^{\mathsf T}=-\sigma_y$; so $2\sigma_{E_j}-\rho_B=\tfrac12s_j\sigma_j=\int_\Lambda(2p(0|j,\lambda)-1)\,\sigma_\lambda\,\mu(d\lambda)$ for each $j$, and
\begin{equation}
3=\sum_js_j\Tr\bigl(\sigma_j(2\sigma_{E_j}-\rho_B)\bigr)=\int_\Lambda\Tr\bigl(\sigma_\lambda\,\vec c_\lambda\cdot\vec\sigma\bigr)\,\mu(d\lambda)\le\sqrt3\int_\Lambda\Tr\sigma_\lambda\,\mu(d\lambda)=\sqrt3,
\end{equation}
with $\vec c_\lambda:=(s_j(2p(0|j,\lambda)-1))_{j\in\{x,y,z\}}\in[-1,1]^3$, since $\vec c_\lambda\cdot\vec\sigma$ has largest eigenvalue $|\vec c_\lambda|\le\sqrt3$, $\sigma_\lambda\succeq0$ and $\int_\Lambda\Tr\sigma_\lambda\,\mu(d\lambda)=\Tr\rho_B=1$; which is impossible. So for this $\effect_B$ the forward direction fails. The reverse direction holds for every $\effect_B$: the only property of $\effect_B$ it uses is that positive semidefinite operators lie in $\effect_B^\circ$, which holds for every effect set.
\end{example}

\section{Conclusions}\label{sec:conclusions}
We have defined two notions of separability relative to an experiment, one for restricted preparations and one for restricted measurements, and placed them on either side of separability itself. Restricting the preparations can only shrink the separable set, and does so as soon as a preparable set is a proper closed convex subset of the state space; restricting the measurements can only enlarge it, and does so, for closed convex effect sets on systems of dimension at least two, as soon as one of them is genuinely restrictive. Outer separability keeps the operational content of separability for the available measurements: it supplies local hidden variable models for the behaviours they generate, and local hidden state models for the assemblages they generate when the steered party is unrestricted.

For the steering question the enlargement is exactly right under two hypotheses: when the effect set of the steering party is generated by finitely many effects that are linearly independent together with the identity and the steered party is unrestricted, the states with local hidden state models for every family of measurements from the effect set are exactly the outer separable states; for two effects the genericity assumption is not needed. Neither hypothesis can be dropped (Examples~\ref{ex:genericity} and~\ref{ex:steered}), and for an effect set not of this form Proposition~\ref{prop:converse-fails} gives hidden state models for a state that is not outer separable.

Several questions remain open. How far can the genericity assumption of Proposition~\ref{prop:osep-lhs-equiv} be weakened for three or more effects? The six effects of Example~\ref{ex:genericity}, which violate the genericity assumption, span the observables together with the identity, while Corollary~\ref{cor:two-povm} needs no genericity assumption for two effects. When the steered party's effect set is restricted, the reverse direction of Proposition~\ref{prop:osep-lhs-equiv} survives and the forward direction can fail (Example~\ref{ex:steered}); does every outer separable state then have, for every family of POVMs from $\effect_A$, a local hidden state model whose hidden states are drawn from the trace-one elements of the dual cone of $\effect_B$ in place of the density matrices? Such models give outer separability by the argument of the proof of Proposition~\ref{prop:osep-lhs-equiv} with $\effect_B^\circ$ in place of the positive semidefinite operators; it is the converse, from outer separability to such models, that is open. For effect sets not of the form of Proposition~\ref{prop:osep-lhs-equiv}, what takes the place of the outer separable set? For the depolarised effect set of Proposition~\ref{prop:converse-fails} it is the whole state space, every state admitting a local hidden state model for every family of measurements from it (its proof). Finally, under the hypotheses of Proposition~\ref{prop:osep-lhs-equiv}, does every state with a local hidden variable model for every behaviour generated by measurements from the two effect sets lie in the outer separable set?

\begin{acknowledgments}
In preparing this paper we used Anthropic's Claude language models (Claude Opus 4.6, 4.7, 4.8, 5 and 5.5, and Claude Fable 5 and 5.1), accessed through the Claude Code environment, for assistance with the literature search and its synthesis, with the derivation and checking of the proofs from the statements, with the drafting and revision of the text, and with the writing and running of numerical checks. The definitions and the statements of the propositions and the corollary were formulated by the authors. We directed every step of this work, re-derived each proof in our own step-by-step working notes, checked the concrete claims numerically and, where they reduce to finite algebra, with the Lean~4 proof assistant and its Mathlib library, and reviewed the derivations. We take full responsibility for the content of this paper.

MP acknowledges support from the Niedersächsisches Ministerium für Wissenschaft und Kultur.
\end{acknowledgments}

\bibliography{citations}

\end{document}